\documentclass{article}

\usepackage{arxiv}

\usepackage[utf8]{inputenc} 
\usepackage[T1]{fontenc}    
\usepackage{hyperref}       
\usepackage{url}            
\usepackage{booktabs}       
\usepackage{amsfonts}       
\usepackage{nicefrac}       
\usepackage{microtype}      
\usepackage{lipsum}
\usepackage{graphicx,textcomp}
\graphicspath{ {./images/} }

\title{Attempting to understand the spread of SARS-CoV-2: Associations between the number of cases in regions and selected economic and geographic parameters}

\author{
 Asit Kumar Mishra\\
  Berkeley Education Alliance for Research in Singapore\\
  Singapore 138602 \\
  \texttt{writeto.asit@gmail.com} \\
   \And
 Pawel Wargocki \\
  Department of Civil Engineering\\
  Technical University of Denmark\\
   Lyngby, Denmark \\
  \texttt{paw@byg.dtu.dk} \\
}

\begin{document}
\maketitle
\begin{abstract}
We present in this work some results from analysing the spread of Covid-19 in different countries and regions around the world and the potential relations with climate, geographical location, and GDP. While the situation remains dynamic, we believe this analysis has the potential to uncover certain underlying trends. We primarily intend the results to drive further, more detailed analysis of the relevant data by other researchers that would help us gather a better understanding of the situation, aiding our preparedness. Our observations show that countries in high latitudes, with temperate and/or continental climate, and with reasonably ''open'' economies are the most vulnerable to this outbreak.
 
\end{abstract}


\section{Introduction}
In the last two months, from a relative unknown, Covid-19 has now become an undeniable reality for people across the entire globe, with nearly a third of the whole world population under lockdowns right now. The situation remains dynamic and fluid, with the number of cases and number of deaths doubling almost every seven days now. In lieu of such conditions, we present here some quick analysis performed on the number of cases in different countries and regions, as related to economic and geographic indicators. We understand that the situation is very dynamic and is changing daily, and the analyses presented will need an update nearly every hour; present analyses are based on the number of cases from March 23, 2020. But we still believe that some trends found in our analysis will remain and may lead to some general conclusions.

\section{Methods}
\label{sec:methods}
We based our analysis on publicly available databases to enhance transparency and so that others can further check the conclusions we reached. The following data sources were used:

\begin{itemize}
	\item Country/region wise case numbers~\cite{worldometer}
	\item GDP per capita, current prices, Purchasing power parity; international dollars per capita~\cite{IMF}
	\item K\"{o}ppen-Geiger climate classification for regions of the world~\cite{KGClimate}
	\item Latitude of centroid of countries/regions~\cite{Centroid}
\end{itemize}

We used the K\"{o}ppen-Geiger (KG) climate classification and we made an approximation to the effect that a country is assigned the KG climate type, which covers the major portion of its land area. In the case where countries have close to equal coverage by two different climate types, we have used binary letter combinations (e.g., AB, CD, etc.).

We assumed a region within 5\textdegree S and 5\textdegree N to make up an equatorial belt. The rest of the world was divided into northern and southern hemispheres.

Data analysis and plotting was done in R 3.5.3~\cite{Rcite}.

\section{Results from the analysis}
\label{sec:results}
\begin{figure}[!htb]
	\centering
	\includegraphics[width=0.75\linewidth]{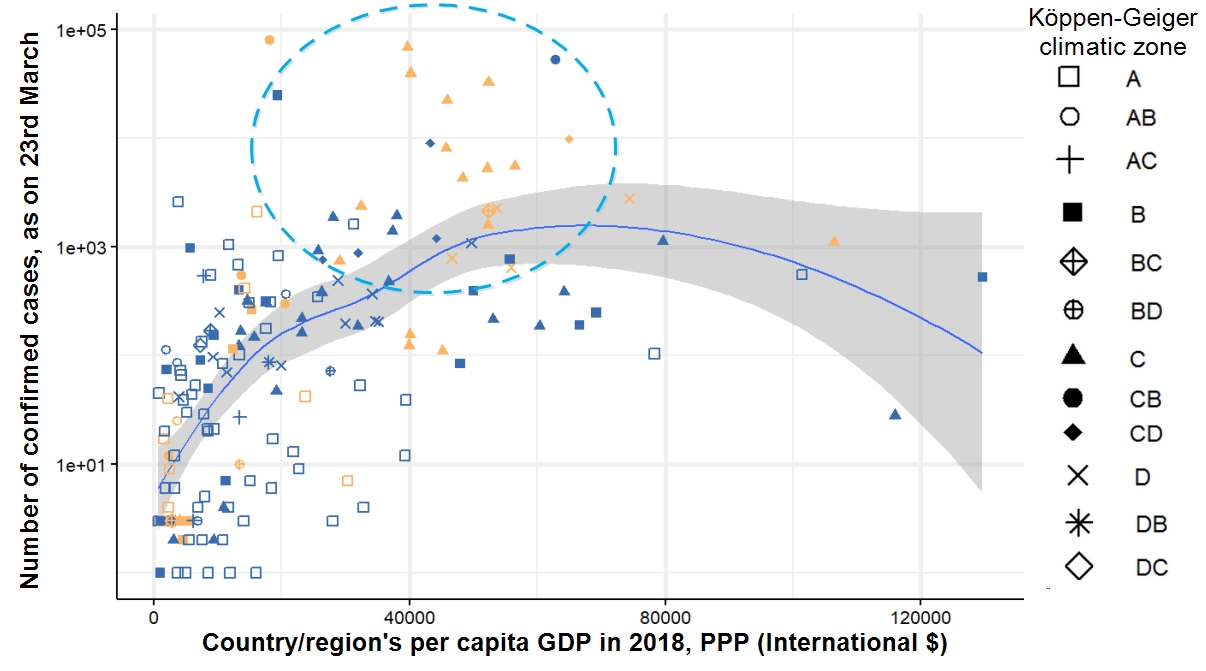}
	\caption{The number of cases on March 23, 2020, and GDP per capita; the markers indicate various climates according to the KG climate classification.}
	\label{fig:CasesVsGDP}
\end{figure}

The number of cases increases with an increase in GDP per capita, but not monotonically (the line in Figure~\ref{fig:CasesVsGDP} uses locally estimated scatterplot smoothing (LOESS) to illustrate the trend). The seeming relation with GDP may be related to the country’s interaction with the international community through trade and other forms of economic exchange. Higher avenues of exchange --- including that of citizens --- is likely to increase the chances of secondary outbreaks. The same idea is also illustrated by dividing the countries into OBOR (golden) and non-OBOR members (blue). As part of China’s OBOR initiative, a country is likely to have much higher economic interaction with China. The gradual increase with per capita GDP seems to peak around 40,000-60,000 International dollars, and has been highlighted by an encasing circle.  

At the two extreme ends, there are two different phenomena. With very high per capita GDP, it is hypothesized that countries are able to better care for their citizens, reducing the number of infections. At the other extreme, countries with low GDP are likely to have minimal interaction with the world economy, providing them a degree of insulation from large outbreaks.

\begin{figure}[!htb]
	\centering
	\includegraphics[width=0.75\linewidth]{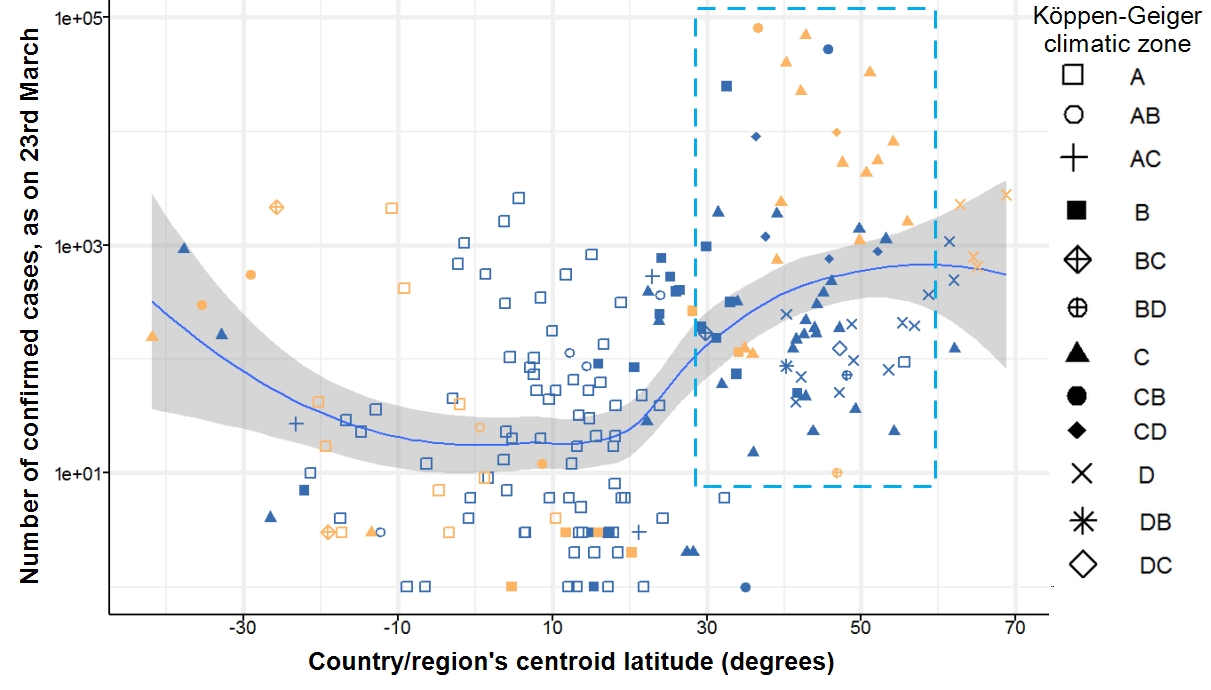}
	\caption{The number of cases on March 23, 2020, and latitude; the markers indicate various climates according to the KG climate classification.}
	\label{fig:CasesVsLatitude}
\end{figure}

The number of cases increases with distance from the equator (the line uses LOESS to illustrate trend), as shown in Figure~\ref{fig:CasesVsLatitude}. At the current time, the outbreak is mostly limited to the Northern Hemisphere, with a sharp rise in numbers seen for latitudes over  30\textdegree N, highlighted by encasing in a rectangle. However, the trend exhibited by data indicates that both over 30\textdegree N latitude and below 30\textdegree S latitude, the chances for more significant outbreaks increase. 

\begin{figure}[!htb]
	\centering
	\includegraphics[width=0.75\linewidth]{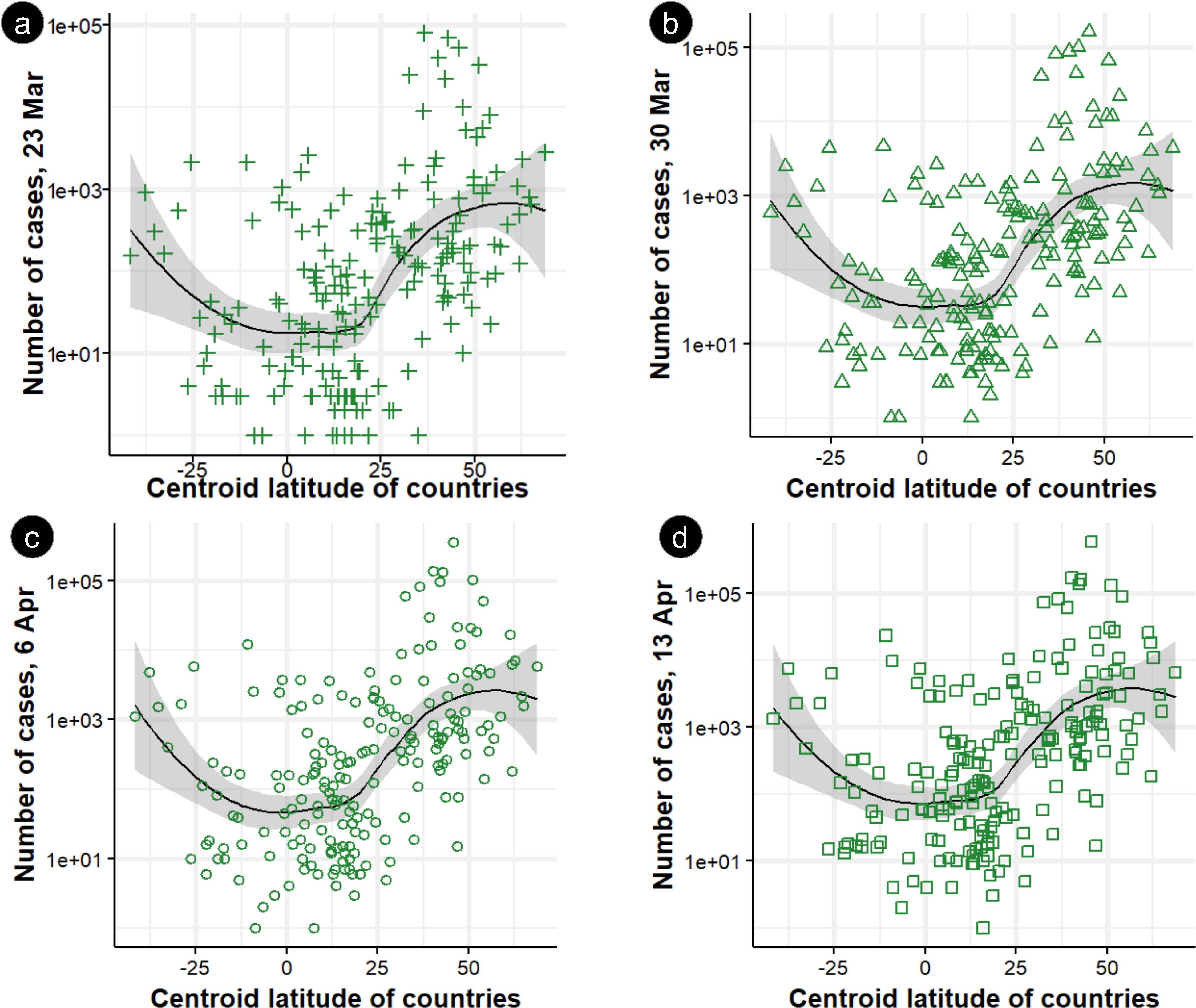}
	\caption{The number of cases, over four weeks, plotted across the latitude of the centroid of the countries, data taken at end of a) March 23, 2020 b) March 30, 2020 c) April 6, 2020 and d) April 13, 2020. Each plot also contains the LOESS fit and the 95\% confidence interval of the fit.}
	\label{fig:CasesVsLatitude4Weeks}
\end{figure}

\begin{figure}[!htb]
	\centering
	\includegraphics[width=0.75\linewidth]{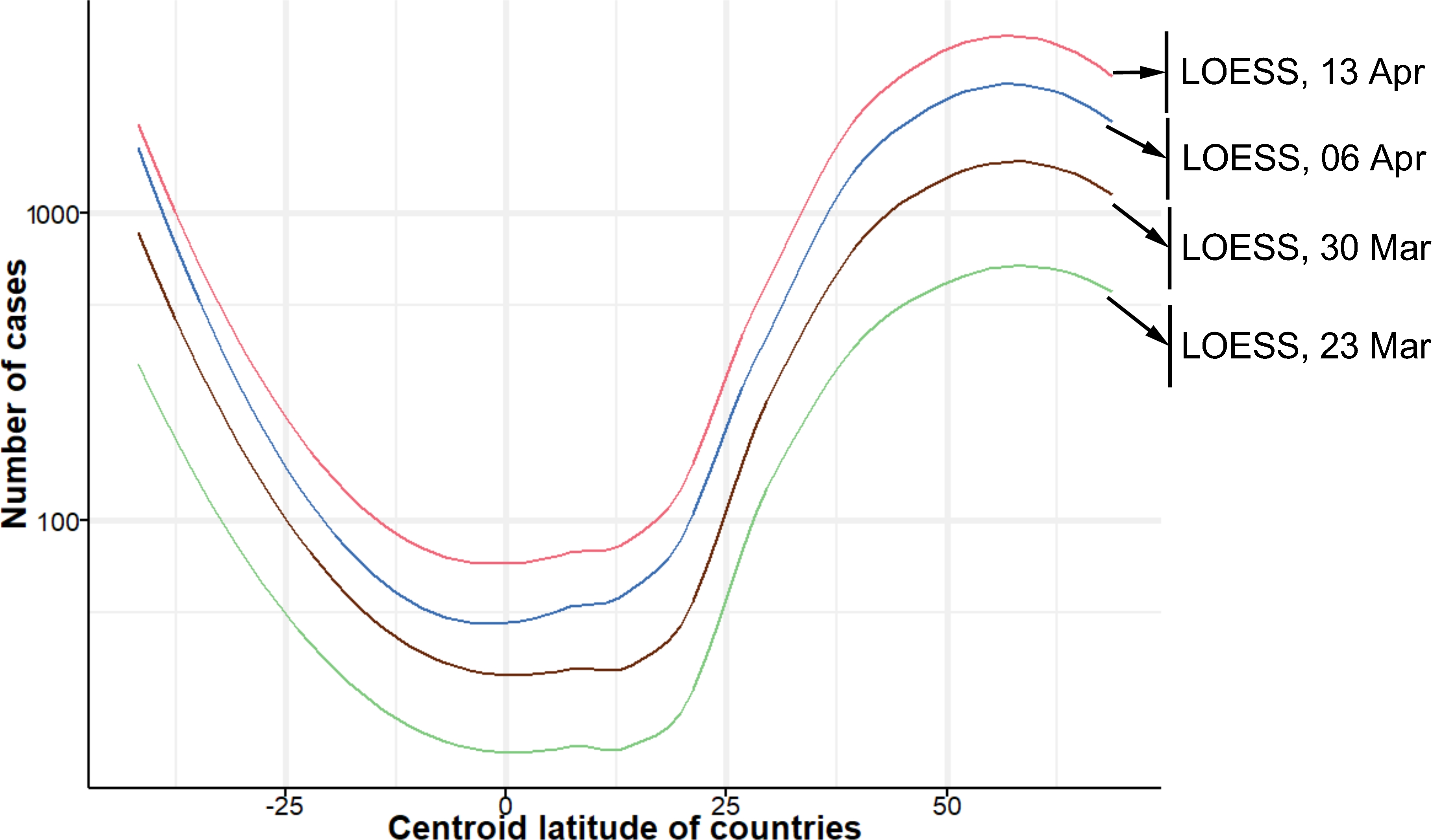}
	\caption{LOESS fits for the number of cases in different countries versus the latitude of the centroid of the country, across four time points, spread over four weeks: March 23, March 30, April 6, and April 13.}
	\label{fig:CasesVsLatitude4WeeksInOne}
\end{figure}

Following the data over a period of 4 weeks, we present in Figure~\ref{fig:CasesVsLatitude4Weeks} the case numbers from March 23, March 30, April 6, and April 13. The LOESS fits for the number of COVID-19 cases in different countries, plotted across the latitude of the centroid of the country, over the same time points, have been presented in Figure~\ref{fig:CasesVsLatitude4WeeksInOne}. From Figures~\ref{fig:CasesVsLatitude4Weeks}~and~\ref{fig:CasesVsLatitude4WeeksInOne}, while the inflation in numbers across the weeks is clear, it is also clear that the trend remains unchanged. In Figure~\ref{fig:CasesVsLatitude4Weeks}, a small tightening of the confidence interval may also be noted with progression of time. Comparatively lower numbers have persisted in latitudes closer to equator (between 25\textdegree~S and 25\textdegree~N). This could be due to the tropical climate (warm and humid conditions), higher UV-C levels, or higher ground level ozone concentrations in tropical regions. However, for now, these are only hypothesises and not enough is known about this novel coronavirus to propose something more concrete. 

\begin{figure}[!htb]
	\centering
	\includegraphics[width=0.75\linewidth]{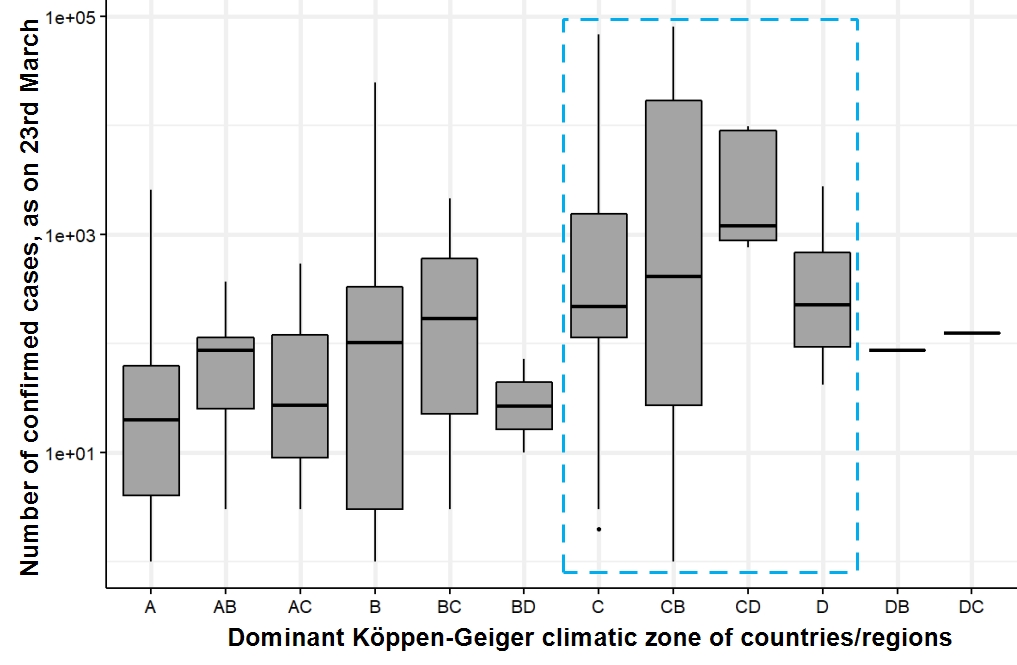}
	\caption{The number of cases on March 23, 2020, represented as boxplots for each climate type, as per KG climate classification.}
	\label{fig:boxplotClimates}
\end{figure}

From Figure~\ref{fig:boxplotClimates}, it is quite apparent that countries with temperate (C) and continental climates (D) have been the worst hit so far, followed by countries with dry climate (B). Also affected are regions with a mix of temperate and dry (CB) or temperate and continental (CD) climates. This could be related to the cool and dry nature of these climates. 

\begin{figure}[!htb]
	\centering
	\includegraphics[width=0.75\linewidth]{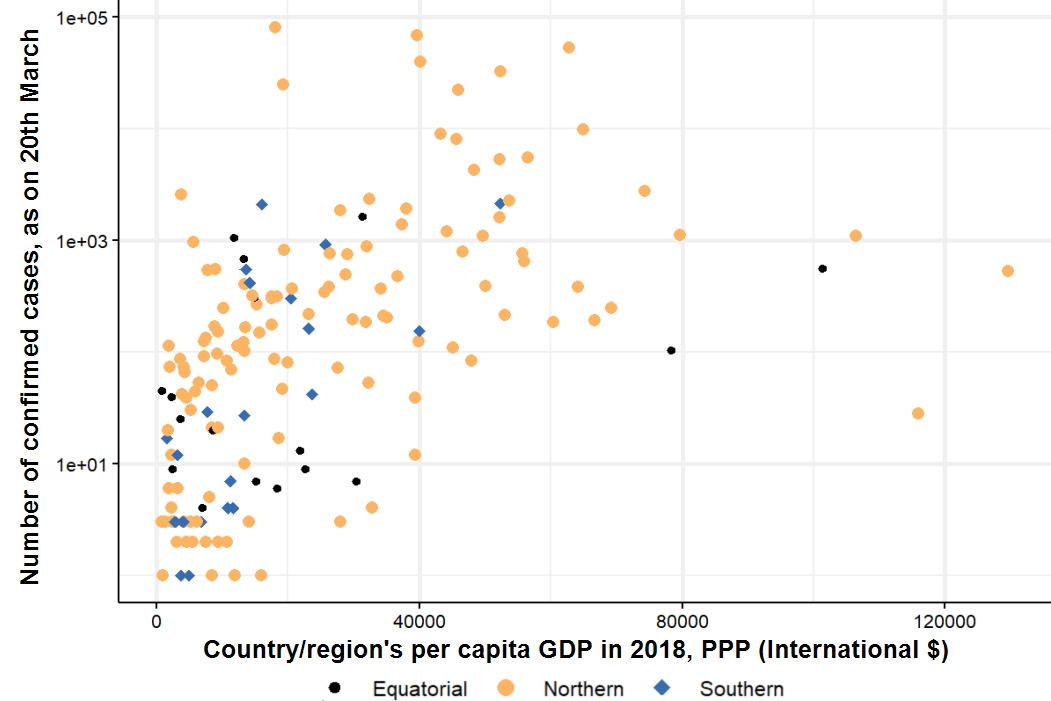}
	\caption{The number of cases on March 23, 2020, and the location of the country/region in equatorial belt, northern hemisphere, or southern hemisphere}
	\label{fig:hemispheres}
\end{figure} 

The outbreak, so far, has been predominantly in the northern hemisphere (Figure~\ref{fig:hemispheres}). But cases in the southern hemisphere are on a gradual rise. Irrespective of GDP variation, cases in regions in the equatorial belt have remained relatively low, so far.

\section{Conclusion}
From our analyses, we drew the following hypotheses:
\begin{enumerate}
	\item Countries with more interaction with world economies have greater vulnerability to outbreaks.
	\item  Countries/regions in high latitudes (over 30\textdegree) in both hemispheres are also particularly vulnerable.
	\item Climate wise, temperate and continental climates are particularly vulnerable --- this may even be true for regional situations in different countries as well.
\end{enumerate}

As the situation remains fluid, we present them only as preliminary hypotheses. We invite other researchers to discuss present analyses, and we hope they will inspire some of them to delve further along these lines.

\bibliographystyle{unsrt}  
\bibliography{references}  

\end{document}